\documentclass[prd,showpacs, twocolumn,amsmath,amssymb]{revtex4}

\usepackage{amssymb}
\usepackage{graphicx}
\usepackage{dcolumn}
\usepackage{bm}
\usepackage{epsfig}
\usepackage{amsfonts}
\usepackage{keyval,graphicx}
\usepackage{textcomp,wasysym}
\usepackage{subfigure}
\usepackage[none]{hyphenat}
\usepackage{xcolor}

\begin{document}
\newcommand{\chisq}[1]{$\chi^{2}_{#1}$}
\newcommand{\etap}{\eta^{\prime}}
\newcommand{\pip}{\pi^{+}}
\newcommand{\pim}{\pi^{-}}
\newcommand{\gam}{\gamma}
\newcommand{\piz}{\pi^{0}}
\newcommand{\rhoz}{\rho^{0}}
\newcommand{\az}{a_{0}(980)}
\newcommand{\fz}{f_{0}(980)}
\newcommand{\pipm}{\pi^{\pm}}
\newcommand{\psip}{\psi(2S)}
\newcommand{\psipp}{\psi^{\prime\prime}}
\newcommand{\jpsi}{J/\psi}
\newcommand{\ar}{\rightarrow}
\newcommand{\GeV}{GeV/$c^2$}
\newcommand{\MeV}{MeV/$c^2$}
\newcommand{\br}[1]{\mathcal{B}(#1)}
\newcommand{\cinst}[2]{$^{\mathrm{#1}}$~#2\par}
\newcommand{\crefi}[1]{$^{\mathrm{#1}}$}
\newcommand{\crefii}[2]{$^{\mathrm{#1,#2}}$}
\newcommand{\crefiii}[3]{$^{\mathrm{#1,#2,#3}}$}
\newcommand{\HRule}{\rule{0.5\linewidth}{0.5mm}}

\title{\boldmath  Partial wave analysis of $\psi(2S) \to  p \bar{p}\eta$}

\author{
{\small
M.~Ablikim$^{1}$, M.~N.~Achasov$^{6}$, O.~Albayrak$^{3}$, D.~J.~Ambrose$^{39}$, F.~F.~An$^{1}$, Q.~An$^{40}$, J.~Z.~Bai$^{1}$, R.~Baldini Ferroli$^{17A}$, Y.~Ban$^{26}$, J.~Becker$^{2}$
% ...
\\
\vspace{0.2cm}
(BESIII Collaboration)\\
\vspace{0.2cm} {\it
$^{1}$ Institute of High Energy Physics, Beijing 100049, People's Republic of China\\
$^{2}$ Bochum Ruhr-University, D-44780 Bochum, Germany\\
$^{3}$ Carnegie Mellon University, Pittsburgh, Pennsylvania 15213, USA\\
% ...
$^{6}$ G.I. Budker Institute of Nuclear Physics SB RAS (BINP), Novosibirsk 630090, Russia\\
% ...
$^{17}$ (A)INFN Laboratori Nazionali di Frascati, I-00044, Frascati, Italy; (B)INFN and University of Perugia, I-06100, Perugia, Italy\\
% ...
$^{26}$ Peking University, Beijing 100871, People's Republic of China\\
% ...
$^{39}$ University of Rochester, Rochester, New York 14627, USA\\
$^{40}$ University of Science and Technology of China, Hefei 230026, People's Republic of China\\
% ...
}}
\vspace{0.4cm} }


\begin{abstract}

  Using a sample of $1.06 \times 10^{8}$ $\psi(2S)$ events collected
  with the BESIII detector at BEPCII, the decay $\psi(2S) \to p
  \bar{p}\eta$ is studied.  A partial wave analysis determines that
  the intermediate state $N(1535)$ with a mass of
  $1524\pm5^{+10}_{-4}$~MeV/$c^2$ and a width of
  $130^{+27+57}_{-24-10}$~MeV/$c^2$ is dominant in the decay; the
  product branching fraction is determined to be $B(\psi(2S) \to
  N(1535)\bar{p})\times B(N(1535)\to p\eta)+c.c.  =
  (5.2\pm0.3^{+3.2}_{-1.2})\times 10^{-5}$.  Furthermore, the
  branching fraction of $\psi(2S) \to \eta p \bar{p}$ is measured to
  be $(6.4\pm0.2\pm0.6)\times 10^{-5}$.

\end{abstract}


\maketitle

\section{Introduction}

Baryon spectroscopy is an important field to understand the internal
structure of hadrons. Within the static quark model, the baryon octet
and decuplet are well described.  About half a century after the
introduction of the quark model, however, a substantial number of
light baryons predicted by the quark model have not been observed
experimentally, which is known as the "missing baryon
problem"~\cite{theory1,theory2}.  One possibility could be that the
missing states simply do not exist, which has lead to the development
of new phenomenological models, eg. the di-quark model~\cite{diquark}.
Alternatively, the coupling of the unobserved states through
conventional production channels could be small, which makes their
observation more difficult.

In addition to fixed target
experiments~\cite{exp1,exp2,exp3,exp4,exp5,exp6,exp7,exp8}, charmonium
decays produced in $e^+e^-$ collisions open a window to hunt for the
missing baryons~\cite{exp9}.  The Beijing
Spectrometer (BES)~\cite{dect10} experiment started a baryon program
about a decade ago with the study of $N(1535)$ and $N(1650)$ in
$\jpsi\rightarrow p\bar{p}\eta$ by partial wave analysis
(PWA)~\cite{pwappbeta} using a sample of 7.8 million $J/\psi$
events. Using 58 million $J/\psi$ events collected at the BESII
detector, a new excited nucleon $N(2065$)~\cite{miss1,miss2} was
observed in $J/\psi\rightarrow p \bar{n}\pi^-$~\cite{nnbpi} and
subsequently confirmed in $J/\psi\rightarrow p\bar{p}\pi^0$
~\cite{pwappbpi0_li}.  BESII also studied $\psi(2S)\rightarrow
p\bar{p} \gamma \gamma$, where both $p\bar{p}\pi^0$ and $p\bar{p}\eta$
were observed, $\psi(2S)\rightarrow p\bar{p}\eta$ for the first time
with a branching fraction of $(5.8\pm1.1\pm0.7)\times 10^{-5}$ . In
both decays, there was weak evidence for a $p\bar{p}$ threshold mass
enhancement but no PWA was performed~\cite{1stetappb}.  Most recently
BESIII reported PWA results of $\psi(2S)\rightarrow p
\bar{p}\pi^0$~\cite{pwappbpi0}, and two new broad excited nucleons,
$N(2300)$ and $N(2570)$, were observed. However, no clear evidence for
$N(2065)$ was found.  Using $24.5\times10^6$ $\psi(2S)$ events,
CLEO-c~\cite{cleoc} reported the analysis of $\psi(2S) \to \gamma
p\bar{p}, \pi^0 p\bar{p}$ and $\eta p\bar{p}$ without considering
interference effects, in which $N(1535)$ and a $p\bar{p}$
enhancement($R_1(2100)$) were investigated in $\psi(2S)$ decay to
$p\bar{p}\eta$.  Those results show that $J/\psi$ and $\psi(2S)$
decays offer a unique place to study baryon spectroscopy.

In this paper, using the 106 million $\psi(2S)$ events taken at the
BESIII detector, a full PWA of the decay $\psi(2S)\rightarrow p
\bar{p}\eta$ is performed.

\section{BESIII Detector and Monte Carlo simulation}

BEPCII ~\cite{dect8} is a double-ring $e^{+}e^{-}$ collider designed
to provide a peak luminosity of $10^{33}$~cm$^{-2}$s$^{-1}$ at a beam
current of $0.93$~A.  The BESIII ~\cite{dect8} detector has a
geometrical acceptance of $93\%$ of $4\pi$ and consists of four main
components: (1) A small-cell, helium-based ($40\%$ He,
$60\%$~C$_{3}$H$_{8}$) Main Drift Chamber (MDC) with $43$ layers
providing an average single-hit resolution of $135$~$\mu$m,
charged-particle momentum resolution in a $1$~T magnetic field of
$0.5\%$ at $1$~GeV/c$^2$, and a $dE/dx$ resolution, which is better
than $6\%$.  (2) A Time-Of-Flight system (TOF) constructed of
$5$-cm-thick plastic scintillators, with $176$ detectors of $2.4$~m
length in two layers in the barrel and $96$ fan-shaped detectors in
the endcaps. The barrel (endcap) time resolution of $80$~ps ($110$~ps)
provides $2\sigma$ K/$\pi$ separation for momenta up to $\sim
1.0$~GeV/$c^2$.  (3) An ElectroMagnetic Calorimeter (EMC) consisting
of $6240$ CsI(Tl) crystals in a cylindrical structure (barrel) and two
endcaps. The energy resolution at $1.0$~GeV is $2.5\%$ ($5\%$) in the
barrel (endcaps), and the position resolution in the barrel (endcaps)
is $6$~mm ($9$~mm) .  (4) The MUon Counter (MUC) consists of
$1000$~m$^{2}$ of Resistive Plate Chambers (RPCs) in nine barrel and
eight endcap layers and provides $2$~cm position resolution.

A GEANT4-based simulation software BOOST~\cite{sim-boost} includes the
geometric and material description of the BESIII detectors, the
detector response and digitization models, as well as the tracking of
the detector running conditions and performance. The production of the
$\psi(2S)$ resonance is simulated by the Monte Carlo (MC) event
generator KKMC~\cite{sim-kkmc}, while the decays are generated by
EvtGen~\cite{sim-evtgen} for known decay modes with branching ratios
being set to the PDG~\cite{pdg2010} world average values, and by
Lundcharm~\cite{sim-lundcharm} for the remaining unknown decays.  The
analysis is performed in the framework of the BESIII Offline Software
System (BOSS)~\cite{ana-boss} which takes care of the detector
calibration, event reconstruction and data storage.


\section{Event selection}
For $\psi(2S) \to p \bar{p} \eta~(\eta \to \gamma \gamma)$, the
topology is quite simple, $p\bar{p}\gamma\gamma$.  Each candidate
event is required to have two good charged tracks reconstructed from the
MDC with total charge zero.  The point of closest approach to the
beamline of each charged track is required to be within $\pm 20$~cm in
the beam direction and $2$~cm in the plane perpendicular to the beam.
Both tracks must have the polar angle $\theta$ in the range of
$|\cos\theta|<0.93$.  The TOF and the specific energy loss dE/dx of a
particle measured in the MDC are combined to calculate particle
identification (PID) probabilities for pion, kaon and proton
hypotheses. The particle type with the highest probability is assigned
to each track. In this analysis, one charged track is required to be
identified as a proton and the other as an anti-proton.

Photon candidates are reconstructed by clustering EMC crystal
energies. For each photon, the minimum energy is $25$~MeV for barrel
showers ($|\cos\theta| < 0.80$) and $50$~MeV for endcap showers
($0.86<|\cos\theta|<0.92$). To exclude showers from charged particles,
the angle between the nearest proton track and the shower must be
greater than $10^{\circ}$, while for the anti-proton the angle has to
be greater than $30^{\circ}$. Timing requirements are used to suppress
electronic noise and energy deposits in the EMC unrelated to the
event.  At least two good photons are required.

\begin{figure}[htbp]
   \vskip -0.1cm
   \centering
   {\includegraphics[width=8cm,height=7cm]{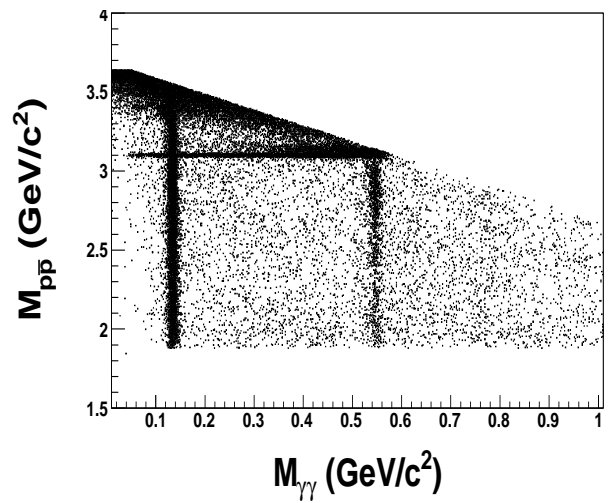}}
   \caption{Scatter plots of $p\overline p$ invariant mass versus $\gamma\gamma$ invariant mass}
   \vskip -0.5cm
   \label{ggppbmass}
\end{figure}

For the candidates remaining, a four-constraint kinematic fit imposing
energy-momentum conservation is made under the $p\bar{p}\gamma\gamma$
hypothesis.  If the number of selected photons is greater than two,
the fit is repeated using all permutations of photons.  The two photon
combination with the minimum fit $\chi^2_{p\bar{p}\gamma\gamma}$ is
selected, and $\chi^2_{p\bar{p}\gamma\gamma}$ is required to be less
than $20$. Because data and MC simulation do not agree well in the low
momentum region, the momenta of the proton and anti-proton are
required to be greater than $300$~MeV/$c$.  Figure~\ref{ggppbmass}
shows the scatter plot of $M_{p\overline p}$ versus $M_{\gamma\gamma}$
for events satisfying the above requirements, where the two vertical
bands correspond to the decays $\psi(2S)\rightarrow p \bar{p}\pi^0$
and $\psi(2S)\rightarrow p \bar{p}\eta$, and the horizontal band
corresponds to the decay $\psi(2S)\rightarrow X +
J/\psi~(J/\psi\rightarrow p\bar{p})$. To remove the background events
from $\psi(2S)\rightarrow \eta J/\psi$ and $\psi(2S) \to \gamma
\chi_{cj}$, $M_{p\bar{p}} < 3.067$~GeV/$c^2$ and $M_{p\bar{p}} <
(3.4$~GeV/$c^2 -0.75 \times M_{\gamma\gamma})$ are required. To select
a clean sample, $M_{\gamma \gamma}$ is required to be in the $\eta$
mass region, $|M_{\gamma \gamma}-M_{\eta}|< 21$~MeV/$c^2$.

\begin{figure*}[htbp]
   \vskip -0.1cm
   \centering{
   {\extracolsep{\fill}}
   \psfig{file=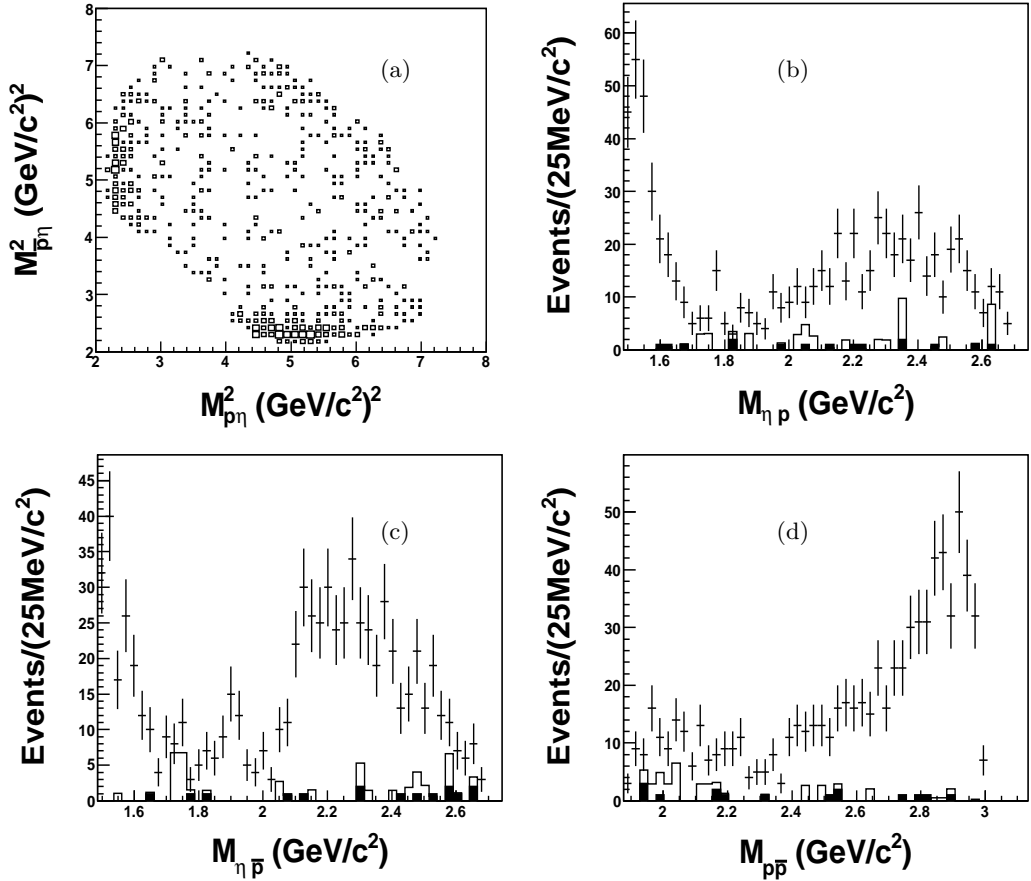,width=14cm,height=12.cm,angle=0}
   \put(-250,310){(a)}
   \put(-100,310){(b)}
   \put(-250,135){(c)}
   \put(-100,135){(d)}
   }
   \vskip -0.5cm
   \caption{ (a) The Dalitz plot of $\psi(2S) \to \eta p\bar{p}$ and
     distributions for (b) $M_{\eta p}$, (c) $M_{\eta \bar{p}}$, and
     (d) $M_{p\bar{p}}$. The crosses represent the data, the blank
     histograms show background events from continuum data, and the
     shaded histograms represent the background events from $\eta$
     sidebands.}
   \label{dalitz}
\end{figure*}

After the above event selection, 745 candidate events are
selected. The Dalitz plot of $M_{p\eta}$ versus $M_{\bar{p}\eta}$ is
shown in Fig.~\ref{dalitz} (a), where two clusters, corresponding to
the $p\eta$ mass threshold enhancement displayed in Fig.~\ref{dalitz}
(b) and Fig.~\ref{dalitz} (c) are visible.  Both the mass spectra and
the Dalitz plot display an asymmetry for $p\eta$ and $\bar{p}\eta$,
which is mainly caused by different detection efficiencies for the
proton and anti-proton.

To investigate possible background events, the same analysis is
performed on the MC sample of 100 million inclusive $\psi(2S)$ events,
and 11 backgound events are found from the channels, $\psi(2S) \to
\gamma \chi_{cJ}(\chi_{cJ}\to p \bar{p} \pi^{0})$, $\psi(2S) \to
\gamma \chi_{cJ}(\chi_{cJ} \to \gamma J/\psi, J/\psi \to \gamma p
\bar{p})$ and $\psi(2S) \to \gamma \chi_{c0}(\chi_{c0} \to \bar{p}
\Delta^{+}, \Delta^{+} \to p \pi^{0})$, which is compatible with the
number of background events, 14, estimated with $\eta$ sidebands
($|M_{\gamma\gamma}-0.43|<50$~MeV/$c^2$ and
$|M_{\gamma\gamma}-0.65|<50$~MeV/$c^2$).  Additionally, $42.6~pb^{-1}$
of continuum data taken at $3.65$~GeV/$c^2$ is used for an estimation
of the background from QED processes and 51 background events are
obtained after normalization with the luminosity of the continuum data
and $\psi(2S)$ data.  The background events from $\eta$ sidebands and
the continuum data will be considered in the PWA of $\psi(2S) \to
p\bar{p}\eta$.

\section{Partial Wave Analysis}

The two-body decay amplitudes in the sequential decay process
$\psi(2S) \to N^* \bar{p}, N^*\to \eta p$ (the charge-conjugate
reaction is always implied unless explicitly mentioned) are
constructed using the relativistic covariant tensor amplitude
formalism ~\cite{fdc01}, and the maximum likelihood method is used in
the PWA~\cite{pwappbpi0_li}.
In $\psi(2S) \to N_X \bar{p}, N_X \to \eta p$, $A_j$ is described as
\begin{equation}
A_j=A^j_{prod-X}(BW)_{X}A_{decay-X};,\end{equation}
where $A^j_{prod-X}$ is the amplitude, describing the production of
the intermediate resonance $N_X$, $BW_X$ is the Breit-Weigner
propagator of $N_X$, and $A_{decay-X}$ is the decay amplitude of
$N_X$.  The total differential cross section $\frac{d\sigma}{d\Phi}$
is
\begin{equation}
\frac{d\sigma}{d\Phi} = |\sum_j c_j A_j + F_{phsp}|^2,
\end{equation}
where $F_{phsp}$ denotes the non-resonant contribution described by a
interfering phase space term.
The probability to observe the event characterized by the measurement
$\xi$ is

\begin{equation}
P(\xi)=\frac{\omega(\xi)\epsilon(\xi)}{\int d \xi \omega(\xi)
\epsilon (\xi)},
\end{equation}
where $\omega(\xi)\equiv\frac{d\sigma}{d\Phi}$ and $\epsilon(\xi)$ is
the detection efficiency.  $\int d \xi \omega(\xi)\epsilon (\xi)$ is
the normalization integral calculated from the exclusive Monte Carlo
sample. The joint probability density for observing $n$ events in the
data sample is

\begin{equation}
 \mathcal{L}
=P(\xi_{1},\xi_{2},...,\xi_{n}) = \prod_{i=1}^{n}P(\xi_{i})
=\prod_{i=1}^{n}\frac{\omega(\xi_i)\epsilon(\xi_i)}{\int d \xi
\omega(\xi)\epsilon(\xi)},
\end{equation}

Rather than maximizing the likelihood function $ln(\mathcal{L})$,
$S=-ln\mathcal{L}$ is minimized to obtain $c_j$ parameters, as well as
the masses and widths of the resonances

\begin{equation}
 -ln\mathcal{L}
 =-\sum^{n}_{i=1}ln(\frac{\omega(\xi_i)}{\int d \xi
\omega(\xi)\epsilon(\xi)})-\sum^{n}_{i=1}ln\epsilon(\xi_i),
\end{equation}

For a given data set, the second term is a constant and has no impact on the determination
of the parameters of the amplitudes or on the relative changes of $\mathcal{S}$ values.
So, for the fitting, $-ln\mathcal{L}$ is defined as

\begin{equation}
 -ln\mathcal{L}
 \equiv -\sum^{n}_{i=1}ln(\frac{\omega(\xi_i)}{\int d \xi
\omega(\xi)\epsilon(\xi)}),
\end{equation}

The contribution of non-$\eta$ events and QED processes can be estimated with $\eta$ sidebands and continuum data.
In the log-likelihood calculation, the likelihood value of $\eta$ sidebands and continuum data events are given negative weights,
and are removed from data, since the log-likelihood value of data is the sum of the log-likelihood
values of signal and background events

\begin{eqnarray}
 S&=&-[(-ln(\mathcal{L}))_{data}-(-ln(\mathcal{L}))_{bg}]  
\end{eqnarray}

The free parameters are optimized by
FUMILI~\cite{fumili}. 
In the minimization procedure, a change in log-likelihood of $0.5$
represents one standard deviation for each parameter.

In the analysis, the following two Breit-Wigner formulas are
used to describe the resonance.  One form has a width which is
independent of the energy of the intermediate state

\begin{equation}
    BW(s)=\frac{1}{M^{2}_{N^{*}}-s-iM_{N^{*}}\Gamma_{N^{*}}},
\end{equation}
where $s$ is the invariant mass-squared.
For $N(1535)$ with its mass close to the threshold of its dominant
decay channel $N\eta$, the approximation of a constant width is not
very good. Thus a phase space dependent width for $N(1535)$ is also
used

\begin{equation}
    BW(s)=\frac{1}{M^{2}_{N^{*}}-s-iM_{N^{*}}\Gamma_{N^{*}}(s)}.
\end{equation}
The phase space dependent
widths can be written as~\cite{BW01}
\begin{equation}\label{eq_width_dependent_BW}
    \Gamma_{N^{*}}(s)=\Gamma_{N^{*}}^{0}(0.5\frac{\rho_{\pi N}(s)}{\rho_{\pi N}(M^{2}_{N^{*}})}+
0.5\frac{\rho_{\eta N}(s)}{\rho_{\eta N}(M^{2}_{N^{*}})}),
\end{equation}
where $\rho_{\pi N}$ and $\rho_{\eta N}$ are the phase space factors
for $\pi N$ and $\eta N$ final states, respectively,

\begin{eqnarray}
    \rho_{XN}(s)&=& \frac{2q_{XN}(s)}{\sqrt{s}} \nonumber \\
                &=& \frac{\sqrt{(s-(M_{N}+M_{X})^{2})(s-(M_{N}-M_{X})^{2})}}{s}, \nonumber \\
\end{eqnarray}
where $X$ is $\pi$ or $\eta$, and $q_{XN}(s)$ is the momentum of $X$
in the center-of-mass system of $XN$.

\begin{figure*}[htbp]
  \vskip -0.1cm \centering{ {\extracolsep{\fill}}
    \psfig{file=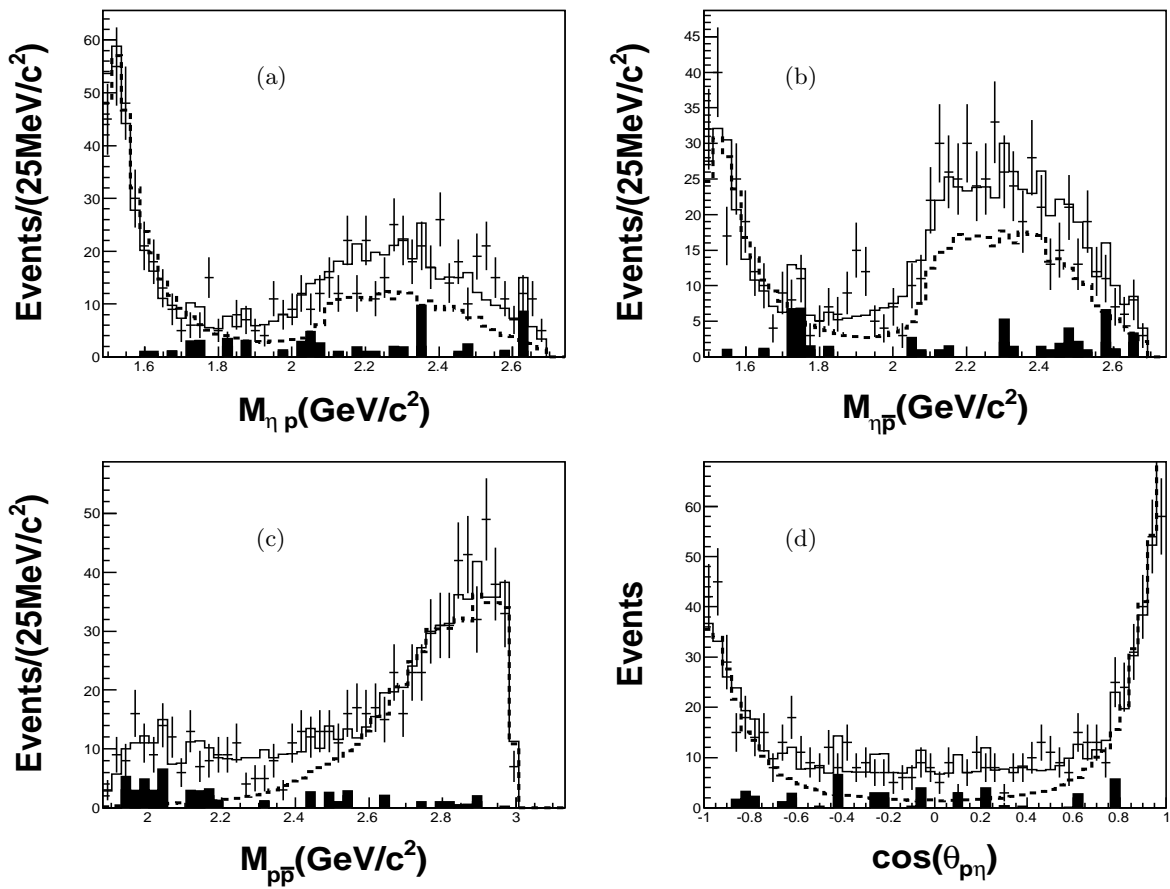,width=16cm,height=12cm,angle=0}
    \put(-350,310){(a)} \put(-150,310){(b)} \put(-350,135){(c)}
    \put(-150,135){(d)} } \vskip -0.5cm
  \caption{Distributions of (a) $M_{p\eta}$, (b) $M_{\bar{p}\eta}$,
    (c) $M_{p\bar{p}}$ and (d) the angle between $p\eta$ in the
    $p\bar{p}$ system. The crosses are for data, the blank histograms
    for PWA projections, the dashed lines for the contribution of
    $N(1535)$ and the shaded histograms for the background events from
    $\eta$ sidebands and continuum data.}
   \label{pwaprojection}
\end{figure*}

\section{systematic errors}
The systematic error sources and their corresponding contributions to the
measurement of mass, width and branching fractions are discussed below.


\begin{itemize}
\item To investigate the impact on the PWA results from other possible
  components, the analysis is also performed including other possible
  $N^*$ states (eg. $N(1520)$, $N(1650)$, $N(1700)$, $N(1710)$,
  $N(1720)$, $N(1895)$ and $N(1900)$); the changes of the mass, width,
  and observed number of $N(1535)\rightarrow p\eta$ events are taken
  as the systematic errors by summing them in quadrature.

\item In the analysis, the background level is quite low, and the events
  from $\eta$ sidebands and continuum data are considered in the
  PWA. To estimate the uncertainty, the background events from $\eta$
  sidebands arew varied by $\pm 50\%$, and the biggest change of the
  results is assigned as the systematic error.

\item In Eq.~(\ref{eq_width_dependent_BW}), the weight of the phase
  space factors for both $\eta N$ and $\pi N$ is set to be 0.5.  The
  change of the results due to the variation of the weights in the
  range of $0\sim1$ is taken as the systematic error.

\item The MDC tracking efficiency was studied with the clean sample of
  $J/\psi \to p \bar{p} \pi^{+} \pi^{-}$ events, as described in
  Ref~\cite{err22}.  The difference between data and MC is less than
  $2\%$ per charged track.  Here, $4\%$ is taken as systematic error
  for the proton and anti-proton.

\item According to the particle identification efficiency study in
  Ref~\cite{err22}, the difference of the particle identification
  efficiencies between MC simulation and data is around 2\% for each
  charged track.  In this study, the two charged tracks are required
  to be identified as $p$ and $\bar{p}$, so $4\%$ is taken as its
  systematic error from this source.

\item The systematic error from the photon detection efficiency has
  been studied using $J/\psi \to\rho^0\pi^0$ events in
  Ref.~\cite{pheff}. The result indicates that the difference between
  data and MC simulation is about 1\% for each photon.  For the decay
  mode analyzed in this paper, 2\% is taken as systematic error from
  two photons in the final states.

\item In order to estimate the systematic error of the kinematic fit, a
  clean sample of $J/\psi \to p \bar{p}\pi^0$ is selected.  The
  difference of the efficiency between data and MC with and without
  using the four constraint kinematic fit, $7\%$, is taken as the
  systematic error.

\item The number of $\psi(2S)$ events, $(1.06\pm 0.86)\times
  10^{8}$~\cite{err23}, was determined from inclusive hadrons, and the
  systematic uncertainty is $0.82\%$.

\end{itemize}

Table ~\ref{tab:pwaerrors} summarizes the systematic error
contributions from different sources for the measurements of mass,
width and branching fractions, and the total is the sum of them in
quadrature.

\begin{table*}[htbp]
\centering
\renewcommand{\arraystretch}{1.5}
\caption{Summary of the systematic error contributions from different sources.}
\begin{tabular*}{0.75\textwidth}{l@{\extracolsep{\fill}}cccc}
  \hline\hline
  Source                & $\Delta$~M(MeV/$c^2$)         &$\Delta~\Gamma$(MeV/$c^2$)    &$\Delta$~B(\%) \\\hline
  Additional Resonances  &$^{+2}_{-4}$ &$^{+56}_{-0}$          &$^{+59}_{-6}$              \\
  Different BW Formula                 &----         &-----                  &$^{+17}_{-21}$    \\
  Background Uncertainty               &$+10$        &$^{+10}_{-10}$         &$+8$                \\
  MDC Tracking              &---- &---- &$\pm4$              \\
  Photon Detection          &---- &---- &$\pm2$              \\
  Particle ID               &---- &---- &$\pm4$              \\
  Kinematic Fit             &---- &---- &$\pm7$              \\
  The number of $\psi(2S)$ events     &---- &---- &$\pm0.82$              \\ \hline

  Total                                &$^{+10}_{-4}$& $^{+57}_{-10}$         &$^{+63}_{-23}$   \\\hline\hline
\end{tabular*}
\label{tab:pwaerrors}

\end{table*}

\section{Results}
The PWA results, including the invariant mass spectra of $p \bar{p}$,
$\eta p$, $\eta \bar{p}$ and angular distributions, are shown as
histograms in Fig.~\ref{pwaprojection} and are consistent with the
data.  The best solution indicates that $N(1535)$ combined with an
interfering phase space is sufficient to describe the data.  We
observe $527\pm 27$ $N(1535)\ar p\eta$ events with a mass $M=
(1524\pm5^{+10}_{-4})$~MeV/$c^2$, a width
$\Gamma=130^{+27+57}_{-24-10}$~MeV/$c^2$, and a statistical
significance larger than $10\sigma$. Here, the first error is
statistical and the second is systematic. The contributions of
$N(1535)$ and phase space are $70.8\%$ and $61.0\%$.  To
determine the detection efficiency of $\psi(2S)\rightarrow
N(1535)\bar{p}$, the MC events are generated in accordance with the
PWA amplitudes for $\psi(2S)\rightarrow N(1535)\bar{p}$. With the
detection efficiency of 24.1\%, the product branching fraction of
$\psi(2S)\to N(1535)\bar{p} ~(N(1535)\to p\eta)$ is calculated to be

\begin{eqnarray}
&&B(\psi(2S)\to N(1535)\bar{p}~(N(1535)\to p\eta))\nonumber  \nonumber \\
&=& \frac{N_{obs}}{\varepsilon\cdot N_{\psi(2S)}\cdot B(\eta \rightarrow \gamma \gamma)}
=(5.2\pm0.3^{+3.2}_{-1.2})\times 10^{-5},\nonumber \\
\end{eqnarray}
where the number of $\psi(2S)$ events, $N_{\psi(2S)}$, is $(1.06\pm
0.86)\times 10^{8}$ determined from $\psi(2S)$ inclusive
decays~\cite{err23}; $B(\eta \to \gamma \gamma)$ is the world average
value~\cite{pdg2012}, and the first error is statistical and the second
systematic.

To investigate the $p\bar{p}$ mass enhancement observed at
BESII~\cite{1stetappb} and CLEO-c~\cite{cleoc}, which did not use a
PWA, a scan for an additional $1^{--}$ resonance described by a
Breit-Wigner function is performed. The widths used are
$50$~MeV/$c^2$, $100$~MeV/$c^2$, $200$~MeV/$c^2$, $300$~MeV/$c^2$,
$400$~MeV/$c^2$, $500$~MeV/$c^2$ and $600$~MeV/$c^2$. The mass is
allowed to vary from $1900$~MeV/$c^2$ to $3000$~MeV/$c^2$ with steps
of $2$~MeV/$c^2$.  There is no evidence for a $p\bar{p}$ resonance in
this region, indicating that the threshold enhancement can be
explained by interference between the $N(1535)$ and phase space.

Subtracting the 51 and 15 background events from QED processes and
from $\eta$ sidebands, respectively, the number of $\psi(2S) \to p
\bar{p}\eta$ events is calculated to be $679\pm 26$. In addition to
the contribution from $N(1535)$, the contribution from the phase space
events is taken into account in the determination of the detection
efficiency according to the PWA results.  With the detection
efficiency of $25.6\%$, the branching fraction of $\psi(2S)
\rightarrow p \bar{p}\eta$ is measured to be

\begin{eqnarray}
B(\psi(2S) \to \eta p \bar{p})&=& (6.4\pm0.2\pm0.6)\times 10^{-5}
\end{eqnarray}

\section{Summary}

Based on $1.06\times 10^{8}~\psi(2S)$ events collected with BESIII
detector, a full PWA on the $745~\psi(2S) \to p \bar{p}\eta$
candidates is performed, and the results indicate that the dominant
contribution is from $\psi(2S) \to N(1535)\bar{p}$. The mass and width
of $N(1535)$ are determined to be $1524\pm5^{+10}_{-4}$~MeV/$c^2$ and
$130^{+27+57}_{-24-10}$~MeV/$c^2$, respectively, which are consistent
with those from previous measurements listed in the PDG
~\cite{pdg2012}.  The product of the branching fractions is calculated
to be $B(\psi(2S)\to N(1535)\bar{p})\times B(N(1535)\to p\eta)+c.c. =
(5.2\pm0.3^{+3.2}_{-1.2})\times 10^{-5}$.  The $p\bar{p}$ mass
enhancement observed by BESII is investigated, and the statistical
significance of an additional $p\bar{p}$ resonance is less than 3
$\sigma$.

The branching fraction of $\psi(2S) \to \eta p \bar{p}$ is determined
to be $(6.4\pm0.2\pm0.6)\times 10^{-5}$, where the detection
efficiency is determined from MC simulation events generated based on
the PWA results.  Compared with the branching fraction of
$J/\psi\rightarrow p\bar{p}\eta$~\cite{pdg2012},

\begin{eqnarray}
Q_{p\bar{p}\eta}&=&\frac{B(\psi(2S) \to \eta p \bar{p})}{B(J/\psi \to \eta p \bar{p})}=(3.2\pm0.4)\%,
\end{eqnarray}
which improves the BESII measurement~\cite{1stetappb} of
$(2.8\pm0.7)\%$, and indicates that the decay $\psi(2S)\rightarrow
p\bar{p}\eta$ is suppressed compared with the "12\% rule".

\section{Acknowledgments}

The BESIII collaboration thanks the staff of BEPCII and the computing center for their hard efforts. This work is supported in part by the Ministry of Science and Technology of China under Contract No. 2009CB825200; National Natural Science Foundation of China (NSFC) under Contracts Nos. 10625524, 10821063, 10825524, 10835001, 10935007, 11125525, 11235011, 10805053; Joint Funds of the National Natural Science Foundation of China under Contracts Nos. 11079008, 11179007; the Chinese Academy of Sciences (CAS) Large-Scale Scientific Facility Program; CAS under Contracts Nos. KJCX2-YW-N29, KJCX2-YW-N45; 100 Talents Program of CAS; German Research Foundation DFG under Contract No. Collaborative Research Center CRC-1044; Istituto Nazionale di Fisica Nucleare, Italy; Ministry of Development of Turkey under Contract No. DPT2006K-120470; U. S. Department of Energy under Contracts Nos. DE-FG02-04ER41291, DE-FG02-05ER41374, DE-FG02-94ER40823; U.S. National Science Foundation; University of Groningen (RuG) and the Helmholtzzentrum fuer Schwerionenforschung GmbH (GSI), Darmstadt; WCU Program of National Research Foundation of Korea under Contract No. R32-2008-000-10155-0


\end{document}